\documentclass[12pt, a4paper]{article}
\usepackage{cmap}
\usepackage[T2a, T1]{fontenc}
\usepackage[cp1251]{inputenc}
\usepackage[russian]{babel}
\usepackage[dvips]{graphicx}
\begin{document}
\title{Конденсация ДНК Т4 в водно-спиртовых средах}
\author{М.\,О.\,Галлямов, О.\,А.\,Пышкина, В.\,Г.\,Сергеев, И.\,В.\,Яминский}
\date{\emph{Физический и Химический факультеты Московского государственного университета им. М. В. Ломоносова, Москва}}

      \maketitle
      \begin{abstract}
      Процесс компактизации высокомолекулярной ДНК\,Т4 исследован в водно-спиртовой среде методом АСМ. Получены АСМ-изображения компактных глобул, образованных молекулами ДНК в водно-спиртовых средах при 80\% концентрации спирта; обнаружено, что при 40--50\% концентрациях спирта молекулы ДНК формируют частично компактизованные образования, в которых отдельные витки макромолекулы закручиваются в тороидальные структуры. С привлечением методики восстановления истинных геометрических параметров объекта по АСМ-профилю, показано, что в состав глобулы входит одна молекула ДНК. По результатам исследований предложена модель упаковки ДНК в процессе компактизации.
\end{abstract}

      Ранее нами было показано\,[1], что гигантская ДНК Т4 в водно-спир\-товой смеси при концентрации спирта более 50\% (для изопропанола) претерпевает конформационный переход клубок$\to$глобула. При уменьшении концентрации спирта до 40\% наблюдалось частичное разворачивание глобул. Однако, в силу ограничения разрешающей способности используемого метода флуоресцентной микроскопии (классический предел $\lambda/2$), оказалось трудным определить микроструктуру полученных объектов. 
      
      Метод сканирующей зондовой микроскопии (СЗМ) может успешно применяться для исследования конформации и микроструктуры молекул ДНК [2,3]. Однако, традиционная схема приготовления образцов для зондовой микроскопии на воздухе включает процедуру высушивания капли препарата на поверхности твердой подложки. Процесс высушивания является неравновесным и сопровождается трудноконтролируемыми изменениями локальных значений концентрации препарата, что может привести к сложностям при интерпретации результатов. Исключить процедуру высушивания в схеме СЗМ-эксперимента можно при проведении исследований в жидкости (с использованием жидкостной ячейки). 

\section*{Экспериментальная часть}

В качестве объектов исследования использовали ДНК бактериофага\,Т4 (Nippone Gen). При проведении экспериментов использовали бидистиллированную деионизованную воду. Водно-спиртовые среды готовили с использованием изопропанола (химически чистого).

      Исследования конденсации молекул ДНК проводили методом атомно-силовой микроскопии (АСМ) непосредственно в \emph{водно-спиртовой смеси}, с использованием подложек слюды. Свежий скол слюды и зонд АСМ помещали в жидкостную ячейку, которую последовательно заполняли используемыми растворами. Препарат, содержащий молекулы ДНК\,Т4 в водно-спиртовой смеси при 80\% концентрации спирта, получали смешивая один объем раствора ДНК в 0,5 ТВЕ-буфере с четырьмя объемами изопропанола. Препарат молекул ДНК в 40\% изопропаноле получали смешивая два объема раствора ДНК в буфере с тремя объемами изопропанола. В обоих случаях конечная концентрация ДНК в смеси составляла $1 \times 10^{-6}$\,моль. Иммобилизация исследуемых структур на подложке осуществлялась лишь за счет процессов адсорбции из раствора.
      
      Эксперименты проводили в жидкостной ячейке АСМ Nanoscope-III (Digital Instruments, USA) в режиме прерывистого контакта с использованием заостренных кантилеверов из нитрида кремния (Nanoprobe) жесткостью 0,32\,Н/м.

\section*{Результаты и их обсуждение}

Препарат, содержащий молекулы ДНК Т4, в водно-спиртовой смеси при 80\% концентрации изопропанола, вводили в жидкостную ячейку, заполненную предварительно водно-спитровой смесью с той же концентрацией спирта. При этих условиях макромолекулы ДНК осаждались на поверхность слюды в виде компактных глобул (рис. 1а и 2а).

\begin{figure}
\begin{center}
\includegraphics*[width = 1.0 \textwidth]{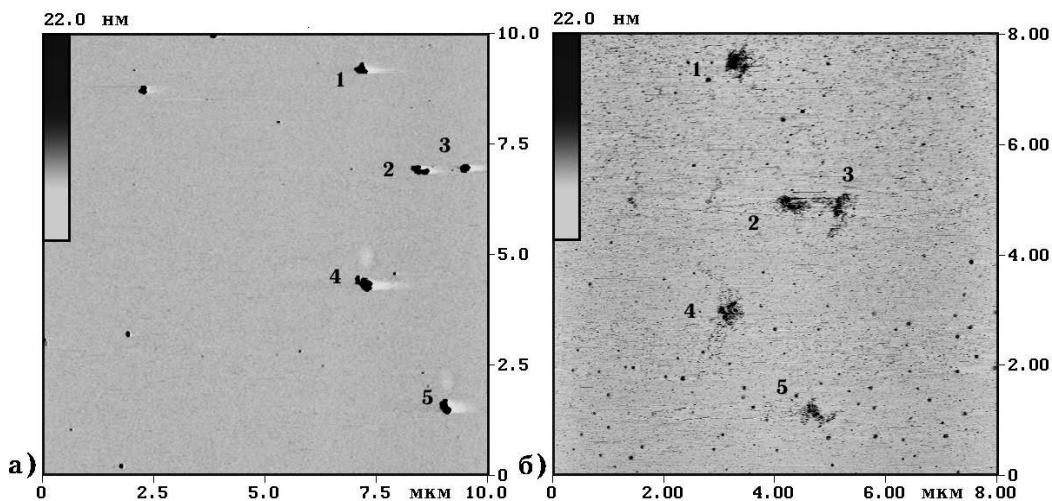}
\caption { a) --- АСМ-изображение компактных глобул, сформированных молекулами ДНК в 80\% изопропаноле, б) --- тот же участок поверхности после понижения концентрации спирта в жидкостной ячейке АСМ до 40--50\%.
}
\end{center}
\end{figure}

\begin{figure}
\begin{center}
\includegraphics*[width = 1.0 \textwidth]{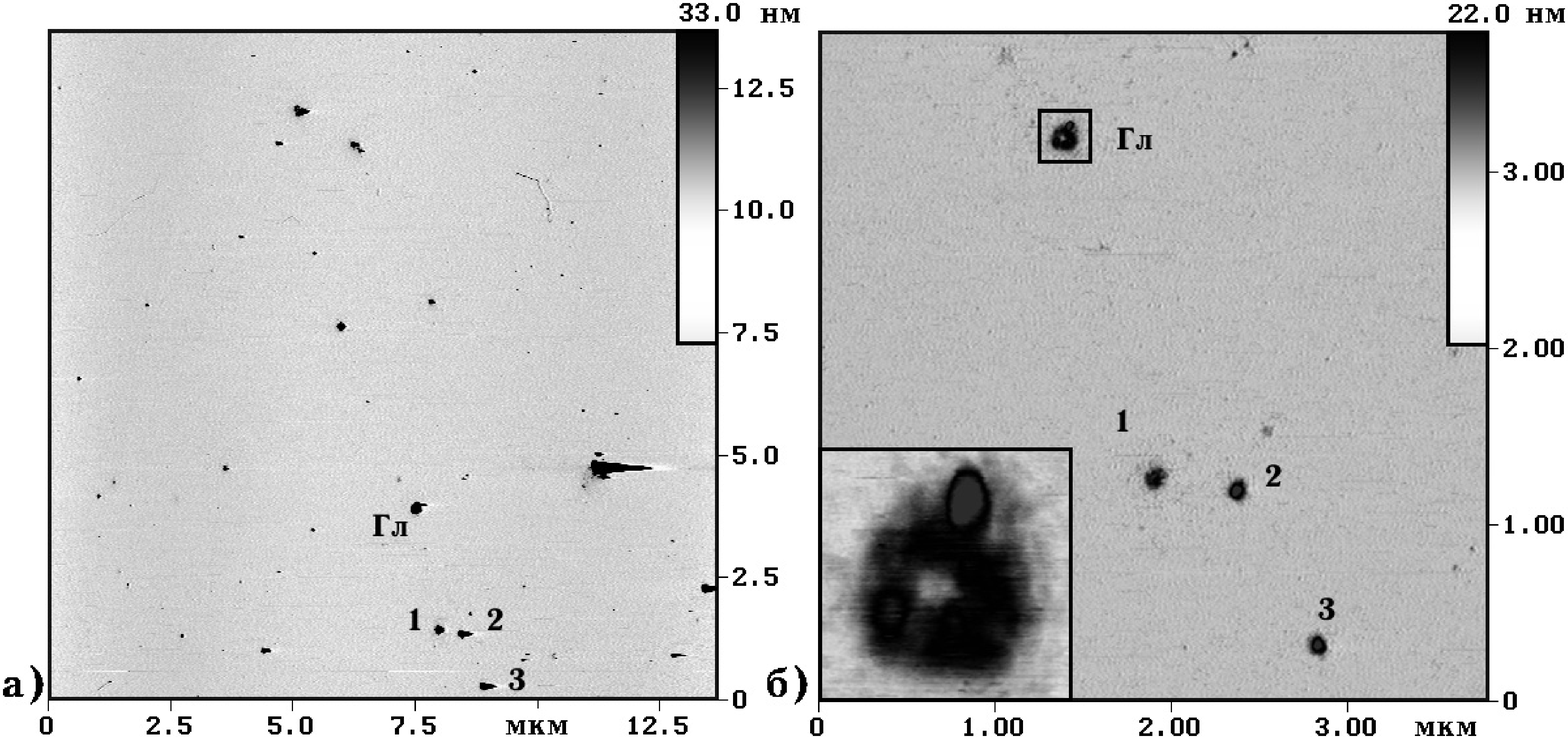}
\caption {a) --- АСМ-изображение компактных глобул, сформированных молекулами ДНК в 80\% изопропаноле, б) --- тот же участок поверхности, после понижения концентрации спирта в жидкостной ячейке АСМ до 40--50\%, врезка --- одна из глобул исследована с большим разрешением, что позволяет визуализовать центральную полость, проявившуюся в результате процесса декомпактизации
}
\end{center}
\end{figure}

      Применение разработанной нами методики восстановления реальных размеров исследуемых объектов (учет \emph{эффекта уширения})\,[4] позволяет определить объем и геометрию глобулярных структур по трем известным параметрам: высоте АСМ-изображения над подложкой, ширине на полувысоте АСМ-профиля (для эллипсоида следует использовать два значения ширины, соответствующие напралениям главных осей) и радиусу кривизны иглы. Мы провели расчеты для двух значений радиуса кривизны иглы: 5 и 10\,нм (по нашим оценкам с применением тест-объектов радиус используемой иглы лежит в указанном интервале), см. табл. 1.

\begin{table}
\begin{center}
\begin{tabular}{|c|c|c|c|c|}
\hline
$R$, нм & $a$, нм & $b$, нм & $c$, нм & $V$, $\times 10^5$\,нм$^3$ \\
\hline
$5$ & $41 \pm 12$ & $55 \pm 17$ & $23 \pm 9$ & $2,5 \pm 1,7$ \\
\hline
$10$ & $37 \pm 12$ & $52 \pm 17$ & $23 \pm 9$ & $2,1 \pm 1,5$ \\
\hline
\end{tabular}
\end{center}
\caption{Геометрические параметры глобул, образованных молекулами ДНК Т4 в результате компактизации спиртом. Расчеты проводились по разработанной нами методике учета \emph{эффекта уширения} [4] для двух значений радиуса кривизны иглы $R$: 5 и 10\,нм, приведенные погрешности являются стандартными отклонениями средних арифметических, рассчитанных путем статистического анализа АСМ-изображений глобул. Обозначения таблицы: 
$a$, $b$ и $c$ --- параметры эллипсоида, описывающего геометрию глобулы; $a$ и $b$ восстановлены по методике учета АСМ-уширения, $c$ --- половина высоты АСМ-изображения глобулы; $V$ --- объем эллипсоида, определенный по формуле: $V_{globule} = 4/3 \pi abc$; для сравнения: объем молекулы ДНК Т4 $V_{DNA} \sim 1,7 \times 10^5 $\,нм$^3$}
\end{table}

      Значения параметров глобулы, определенные с использованием двух граничных значений для радиуса кривизны иглы, фактически, идентичны (с учетом значительной величины стандартного отклонения, что обусловлено статистическим разбросом анализируемых параметров для различных глобул). Восстановленные значения параметров полуосей $a$ и $b$ позволяют сделать вывод, что геометрической формой глобулы является сплюснутый и слегка вытянутый эллипсоид. 
      
      Из анализа таблицы следует, что объем глобулы составляет величину $(2,5 \pm 1,7) \times 10^5$\,нм$^3$, что превышает объем молекулы ДНК\,Т4: $1,7 \times 10^5 $\,нм$^3$ ($V_{DNA} =\pi r^2 L$, $r = 1$\,нм, $L = 55 \times 10^3$\,нм). Т.о. образом, можно предположить, что каждая глобула образованна \emph{одной} молекулой ДНК, находящейся в достаточно плотноупакованном состоянии.
      
      Частично декомпактизованные структуры макромолекул ДНК\,Т4, получаемые при разбавлении 80\% раствора в воде до 40--50\%, исследовали следующим образом. В жидкостную ячейку, содержащую ДНК Т4 в виде глобул в 80\% изопропаноле (20\% воды), добавляли смесь изопропанола с водой (40\% изопропанола + 60\% воды).
      
      Из рис. 2а (отметка «Гл») видно, что при 80\% изопропанола частицы ДНК представляют собой компактные глобулы. При прокачивании через ячейку 40\%-ного раствора изопропанола начинается динамический процесс декомпактизации --- глобулы уменьшаются по высоте и в некоторых из них (рис. 2б и врезка) проявляется центральная полость. Однако оказалось, что дальнейший процесс разворачивания глобул не происходит. Это можно объяснить наличием сил взаимодействия между молекулой и подложкой, которые и препятствуют разворачиванию при уменьшении концентрации спирта. Поэтому макромолекулы ДНК в промежуточном состоянии (между глобулой и клубком) получали следующим образом.
      
      Молекулы ДНК, находящиеся в 40\% изопропаноле, вводили в жидкостную ячейку, а затем концентрацию спирта в ячейке повышали прокачиванием 80\% изопропанола. При повышении концентрации спирта молекулы выпадали на поверхность подложки в частично компактизованном состоянии. Дальнейшая компактизация ДНК, адсорбированных на подложку, не наблюдалась, что, по-видимому, объясняется взаимодействием молекул ДНК со слюдой.
      
      АСМ-изображения частично компактизованных структур приведены на рис.\,3 и 4. Было обнаружено, что начальным процессом компактизации глобул является закручивание отдельных участков макромолекулы ДНК в тороидальные структуры, эти участки, по-видимому, и являются центрами дальнейшей компактизации.
      
\begin{figure}
\begin{center}
\includegraphics*[width = 1.0 \textwidth]{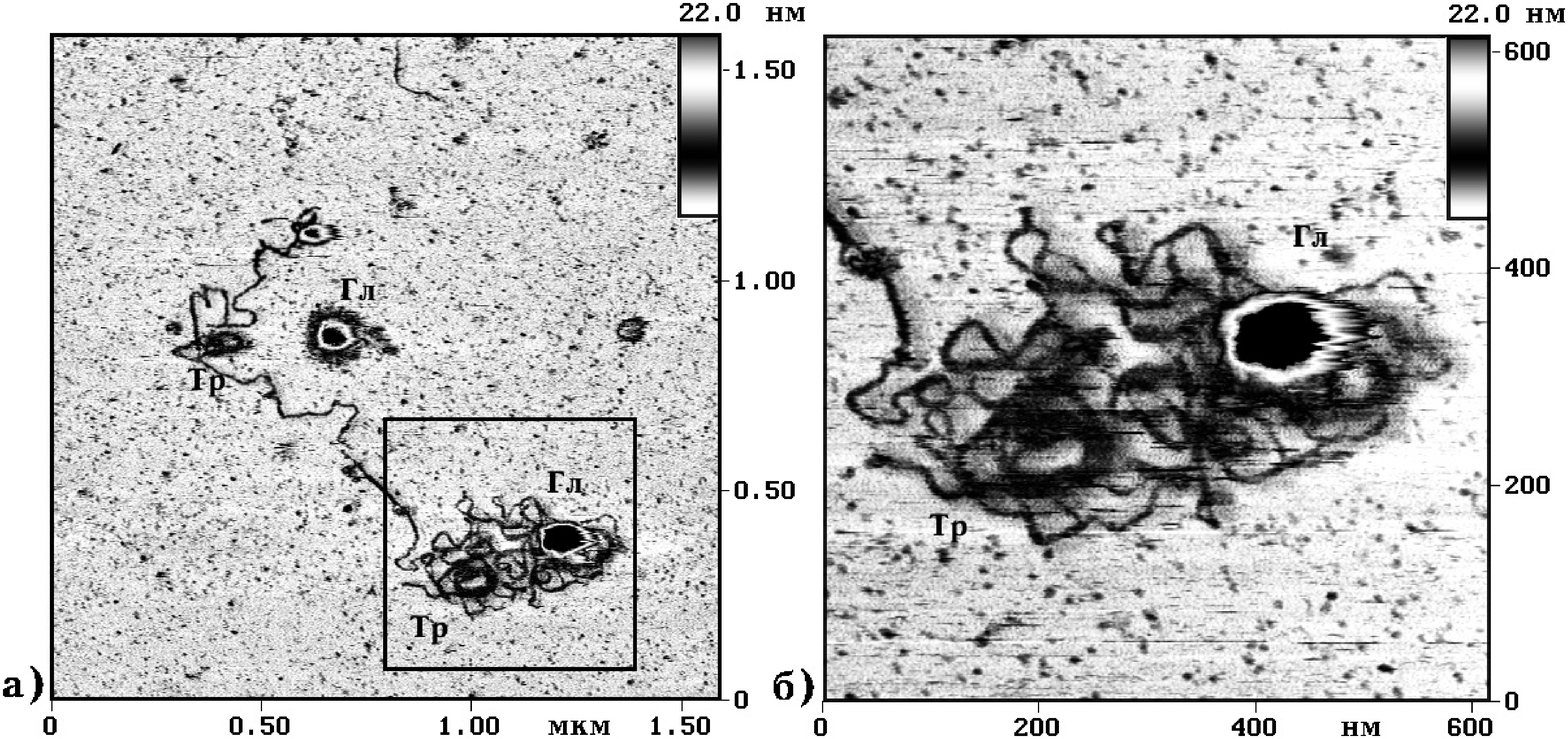}
\caption {Частично компактизованные молекулы ДНК, сформированные и визуализованные в жидкостной ячейке АСМ при повышении концентрации изопропанола (от 40\%), на рисунке отмечены глобулярные («Гл») и тороидальные («Тр») образования
}
\end{center}
\end{figure}

\begin{figure}
\begin{center}
\includegraphics*[width = 1.0 \textwidth]{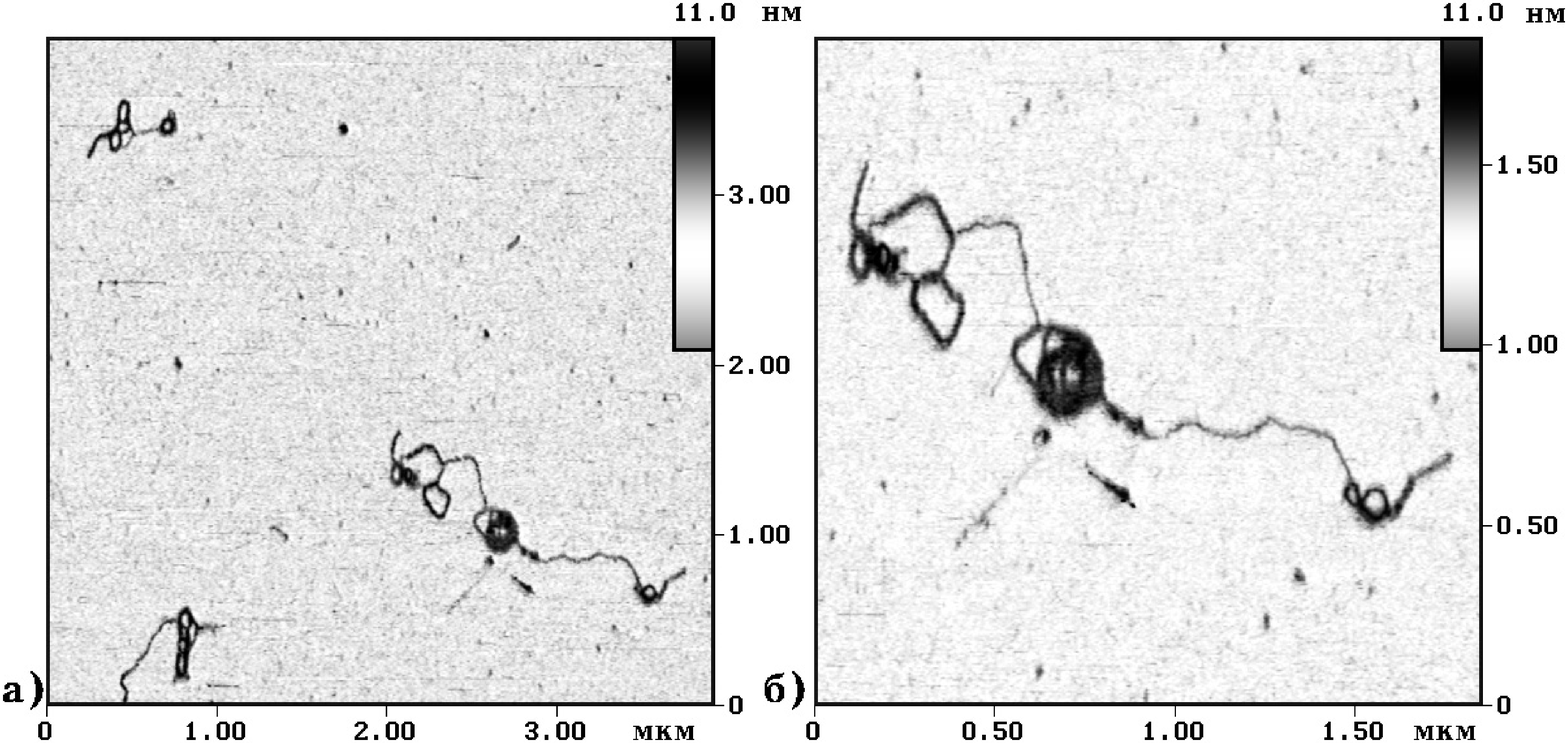}
\caption {Частично компактизованные молекулы ДНК, сформированные и визуализованные в жидкостной ячейке АСМ при повышении концентрации изопропанола (от 40\%); из рисунка видно, что в данных условиях отдельные участки молекулы организуются в тороподобные структуры
}
\end{center}
\end{figure}

      На основании наблюдений можно предположить, что компактные глобулы, приведенные на рис.\,1а и 2а, являются продуктом именно тороидальной компактизации макромолекул ДНК и представляют собой частицы высоко плотности, в которых отдельные участки молекулы закручены в тороидальные структуры. При этом в центре компактной частицы возможно имеется полость: действительно, объем глобулы несколько превышает объем одной молекулы ДНК\,Т4, и частично декомпактизованные глобулы позволяют визуализовать центральную полость (рис.\,2б и врезка).

\begin{figure}
\begin{center}
\includegraphics*[width = 1.0 \textwidth]{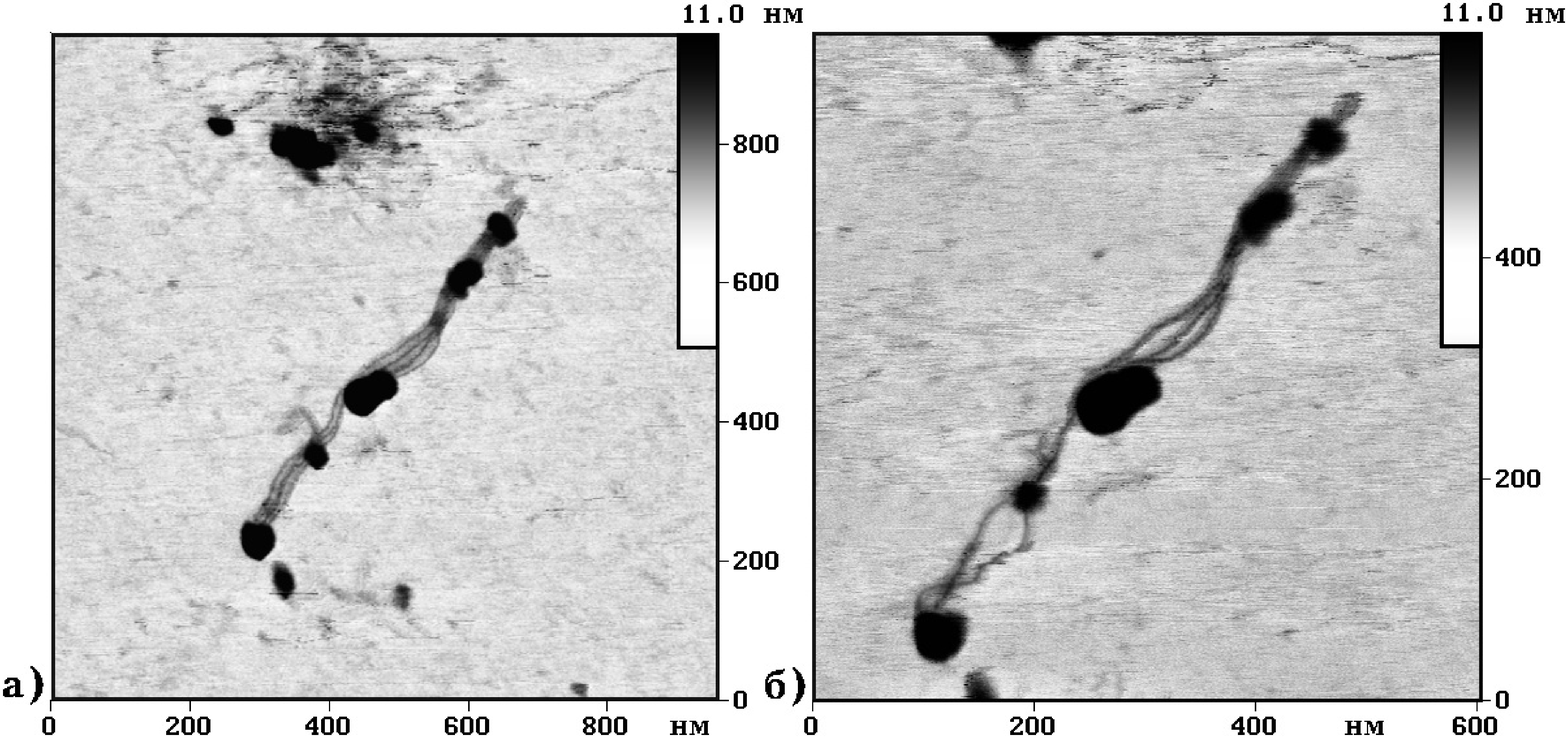}
\caption {а) --- стержневая структура, образованная в результате компактизации молекулы ДНК, визуализована при повышении концентрации спирта (от 40\%) в жидкостной ячейке АСМ, б) --- та же структуры частично ``расплелась'' под воздействием зонда микроскопа после нескольких сканирований
}
\end{center}
\end{figure}

      Следует отметить, что макромолекулы ДНК при повышении концентрации спирта в ячейке (от 40\%) образуют не только тороидальные, но и (реже) стержнеобразные структуры (рис.\,5). Процесс сканирования приводит к тому, что стержневая структура частично «расплетается» под воздействием зонда (см. рис.\,5а и 5б). Наблюдение в ряде случаев компактизованных структур, имеющих морфологию, отличную от тороидальной, может объясняться тем, что в эксперименте исследовались промежуточные стадии процесса компактизации, в которых некоторое количество молекул может находиться в нестационарных (неравновесных) морфологических состояниях.

     \emph{Благодарность}. Работа выполнена при финансовой поддержке Федеральной программы «Университеты России --- фундаментальные исследования» (проект \No 5060) и РФФИ (проект \No 97-03-32778а).

\section*{T4 DNA condensation in water-alcohol media}
\subsection*{M. O. Gallyamov, O. A. Pyshkina, V. G. Sergeyev, I. V. Yaminsky}

The process of compaction of high molecular weight DNA Т4 is investigated directly in a AFM liquid cell. The AFM-images of globules formed by DNA molecules in the result of compaction in water-alcohol environments at high izopropanol concentration (80\%) are received; it is found that at intermediate concentration of izopropanol (40--50\%) the DNA molecules form partially compacted formations in which the separate coils of macromolecules twist in toroidal structures. It is shown using the technique of deconvolution of the AFM-images that the globule include only one closely packed DNA molecule. The model of DNA packing is proposed on the basis of AFM experiment.

\end{document}